\documentclass[conference]{IEEEconf}

\input epsf
\usepackage{graphicx}
\usepackage{multirow}
\usepackage{titlesec}
\usepackage{balance} 
\usepackage{stfloats}
\usepackage{xcolor}
\usepackage{cite}
\usepackage{hyperref}
\usepackage{tcolorbox}
\usepackage{listings}
\usepackage{caption}
\usepackage{subcaption}
\usepackage[font=small,skip=1pt]{caption}
\def\BibTeX{{\rm B\kern-.05em{\sc i\kern-.025em b}\kern-.08em
    T\kern-.1667em\lower.7ex\hbox{E}\kern-.125emX}}
\renewcommand\thesection{\arabic{section}} % arabic numerals for the sections

\renewcommand\thesubsectiondis{\thesection.\arabic{subsection}}% arabic numerals for the subsections

\renewcommand\thesubsubsectiondis{\thesubsectiondis.\arabic{subsubsection}}% arabic numerals for the subsubsections

\begin{document}

\title{\textbf{\Large IDOL: Improved Different Optimization Levels 
Testing for Solidity Compilers\\}}

\author{Lantian Li, Yejian Liang, and Zhongxing Yu$^{*}$\\
	\normalsize Shandong University, Qingdao, China\\
	\normalsize lilantian@mail.sdu.edu.cn, yejianliang@mail.sdu.edu.cn, zhongxing.yu@sdu.edu.cn\\
	\normalsize *corresponding author
}
%+++++++++++++++++++++++++++++++++++++++++++

%\specialpapernotice{(Invited Paper)}

\maketitle
\begin{abstract}
As blockchain technology continues to evolve and mature, smart contracts have become a key driving force behind the digitization and automation of transactions. Smart contracts greatly simplify and refine the traditional business transaction processes, 
% enhance transparency and trust in transactions, 
and thus have had a profound impact on various industries such as finance and supply chain management.
% , and intellectual property protection. 
However, because smart contracts cannot be modified once deployed, any vulnerabilities or design flaws within the contract cannot be easily fixed, potentially leading to significant financial losses or even legal issues. The compiler, as a critical component in the development process, directly affects the quality and security of smart contracts. This paper innovatively proposes a method, known as the Improved Different Optimization Levels (IDOL), for testing the Solidity compiler. The key idea behind IDOL is to perform reverse optimization transformations (\emph{i.e.}, change optimized form into unoptimized form) to generate semantically equivalent variants of the smart contracts under test, aiming to maximize the opportunities to trigger the compiler’s optimization logic.
%This method applies compiler optimization rules to subject the contract under test to equivalent mutations, thereby more easily triggering the compiler’s optimization process, and then performs tests across different optimization levels. 
We conducted a preliminary evaluation of IDOL and three confirmed compiler optimization bugs have been uncovered at the time of writing.

% The IDOL testing method was applied to 160,000 test contracts, and the preliminary results 
% The test results indicate that the IDOL, as a means of assessing the compiler's optimization capabilities, revealed three optimization-related compiler bugs.
\end{abstract}
\IEEEoverridecommandlockouts
\vspace{1.5ex}

% for peerreview papers, inserts a page break and creates the second title.
% Will be ignored for other modes.
\IEEEpeerreviewmaketitle

\section{Introduction}
%The decentralized nature of blockchain technology enhances the transparency, security, and traceability of transactions, bringing groundbreaking changes to the financial sector. 
As blockchain technology continues to evolve and mature, smart contracts have become a key driving force behind the digitization and automation of transactions.
% In this context, smart contracts, as one of the most transformative applications of blockchain technology, play a central role, especially on blockchain platforms such as Ethereum. 
A smart contract is a distributed computing paradigm that can automatically execute, control, or record events and actions between parties~\cite{fsesolidity}.
% enabling trustworthy interactions without the need for intermediaries~\cite{wood2014ethereum}. 
The smart contract development language, such as Solidity, provides an efficient way of implementing the automatic execution of contracts. However, the immutable nature of smart contracts
% , while ensuring the security and transparency of transactions, 
also introduces new challenges. Since smart contracts cannot be modified once deployed, any vulnerabilities within the contract cannot be easily fixed. Thus, ensuring the security of smart contracts is of great importance.

The smart contract compiler is a critical component in the smart contract development process, and its security directly impacts both the quality of the contracts deployed on the blockchain and the overall system security. Thus, systematic testing of the compiler
is of vital importance.
%not only enhances the security of smart contracts, but also serves as a key safeguard for the development of blockchain technology. 
Compiler errors may arise at various stages of the compilation process, such as syntax analysis, semantic analysis, and code generation. Moreover‌, to generate efficient target code, the compiler must also account for the characteristics and optimization requirements of different hardware platforms. As such, ensuring the compiler's quality is an inherently challenging task. In particular, when using a specific program as test input, determining whether the compiler's output is correct can be difficult due to the lack of explicit specifications. That is, the ``test oracle" issue is serious for compiler testing~\cite{barr2014oracle}.

%This issue relates to the ``Test Oracle" problem, which refers to the challenge of defining the correctness of the test output~\cite{barr2014oracle}.

Randomized Differential Testing (RDT)~\cite{groce2007randomized} is an approach that combines randomized testing with differential testing to address the oracle issue. This method generates identical random inputs for multiple systems with different implementations or configurations, and compares their outputs to identify potential errors. Since RDT does not rely on predefined correct outputs, it is well-suited for complex systems that lack explicit specifications, such as compilers and interpreters. Given that many compiler errors are often triggered during the optimization phase~\cite{compileroptimization}, Different Optimization Levels (DOL) testing was proposed as a specialized form of RDT~\cite{chen2016empirical}. This method applies different optimization levels to the same test program on the same compiler, and compares the consistency of the output results to detect potential optimization errors.

To more effectively detect bugs in smart contract compilers, this paper innovatively proposes an Improved Different Optimization Levels (IDOL) testing method to test the Solidity compiler. The key idea behind IDOL is to perform reverse transformations on typical compiler optimizations to generate semantically equivalent variants of the smart contracts under test, aiming to maximize the opportunities to trigger the compiler’s optimization logic.
% , and then performs output consistency testing at different optimization levels. erforming reverse transformations on these typical optimizations
IDOL consists of three major steps.
First, a set of original smart contracts are generated in bulk using Solsmith\footnote{\url{https://github.com/logicseek/Solsmith}}. Then, for each original smart contract, semantically equivalent variants are generated by referencing typical optimization strategies and specific optimization rules of the Solidity compiler. 
% This helps to more easily trigger the compiler's optimization process, thereby activating more optimization behaviors. 
Finally, each variant is compiled and executed under different optimization levels, and the output results are compared. If differences are found in the execution results at different optimization levels, it indicates that there is a bug in at least one optimization path. 
%Preliminary evaluation results show that the IDOL testing method exhibits effective bug detection capabilities in assessing both compiler correctness and optimization performance.
We conducted a preliminary evaluation of IDOL and three confirmed compiler optimization bugs have been uncovered at the time of writing. 
% have uncovered three confirmed compiler optimization bugs at the time of writing.

\vspace{-3pt}

\section{The IDOL  Approach}
\vspace{-2pt}
The \texttt{--optimize-runs} parameter in the Solidity compiler is used to specify the expected execution frequency of each code segment throughout the contract's lifecycle, thereby guiding the optimizer to balance between deployment size and runtime cost. Smaller values typically generate bytecode that is smaller in size but less efficient in execution, making it suitable for low-frequency invocation scenarios. Larger values instead prioritize execution efficiency, making them more appropriate for high-frequency invocation logic. Given the large number of optimization parameters available, to simplify the experiments and cover typical optimization scenarios, we selected three representative configurations for testing: optimization disabled (\texttt{--optimize=false}), optimization enabled with \texttt{runs=1}, and optimization enabled with \texttt{runs=200}.

Unlike the conventional approach which directly applies different optimization levels to the generated test programs, our approach first performs semantically equivalent transformations on the test programs to make it easier to trigger potential errors in the compiler's optimization process. We then proceed with testing under different optimization levels. The equivalent transformations arise from compiler optimizations, currently including loop invariant code motion, loop inversion, 
%loop condition hoisting, 
and several optimization strategies specific to the Solidity compiler. \emph{By performing reverse optimization transformations, we can induce the compiler to execute the corresponding optimization processes without altering the program semantics, thus increasing the likelihood of bug triggering}. Below we introduce two of the considered compiler optimization strategies.

\textbf{Loop-invariant Code Motion.} 
It moves loop-invariant statements or expressions outside the loop body without changing the semantics of the program. The two operations \texttt{x = y + z} and \texttt{x * x} in Figure~\ref{invariantO} are loop invariants and will be moved out of the loop body during compilation optimization, thus transforming into the form of Figure~\ref{invariantM}.

\begin{figure}[!t]
    \centering    
\begin{subfigure}[t]{0.20\textwidth}
        \centering
            \begin{lstlisting}[xleftmargin=1em, xrightmargin=0em]
for (int i = 0; i < n; i++) {
    x = y + z;
    a[i] = 6 * i + x * x;
}
            \end{lstlisting}
        \caption{Original Code}
        \label{invariantO}
    \end{subfigure}
    \hfill
    \begin{subfigure}[t]{0.20\textwidth}
        \centering
            \begin{lstlisting}
x = y + z;
t1 = x * x;
for (int i = 0; i < n; i++) {
    a[i] = 6 * i + t1;
}
            \end{lstlisting}
        \caption{Optimized Code}
        \label{invariantM}
    \end{subfigure}  
    \hfill
    \setlength{\belowcaptionskip}{-10pt}
    \caption{Loop-invariant Code Motion Optimization.}
    \label{fig:invariant}
\end{figure}

\textbf{Loop Inversion.} 
This optimization transformation replaces the \texttt{while} loop with an \texttt{if} block containing a \texttt{do..while} loop. Although the optimized code in Figure~\ref{InversionM} may appear more complex than the original code in Figure~\ref{InversionO}, it may actually run faster. This is because modern CPUs use instruction pipelines and any jump in the code can cause a pipeline stall, which degrades performance. 

Note that the IDOL transformations are reverse, \emph{i.e.}, they convert the optimized form back to the unoptimized form.
\vspace{-2pt}
\begin{figure}[!t]
    \centering    
\begin{subfigure}[t]{0.20\textwidth}
        \centering
            \begin{lstlisting}[xleftmargin=1em, xrightmargin=0em]
int i, a[100];
i = 0;
while (i < 100) {
    a[i] = 0;
    i++;
}

            \end{lstlisting}
        \caption{Original Code}
        \label{InversionO}
    \end{subfigure}
    \hfill
    \begin{subfigure}[t]{0.20\textwidth}
        \centering
            \begin{lstlisting}
int i, a[100];
i = 0;
if (i < 100) {
    do {
        a[i] = 0;
        i++;
    } while (i < 100);
}
            \end{lstlisting}
        \caption{Optimized Code}
        \label{InversionM}
    \end{subfigure}  
    \hfill
    \setlength{\belowcaptionskip}{-15pt}
    \caption{Loop Inversion Optimization.}
    \label{fig:inversion}
\end{figure}

\section{Preliminary Evaluations and Results}
\begin{figure}[htbp]
\centering
\includegraphics[width=0.43\textwidth]{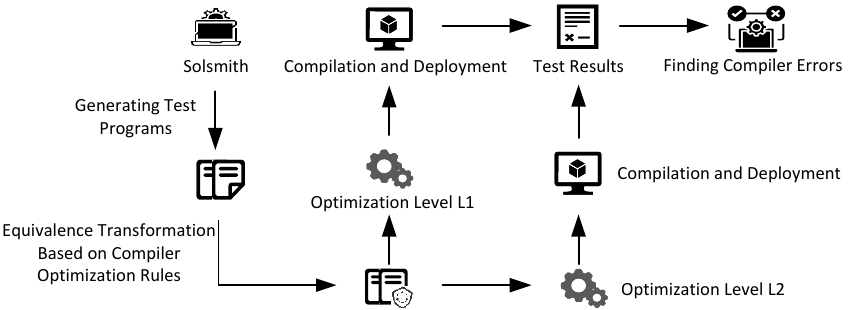}
\setlength{\belowcaptionskip}{-6pt}
\caption{The IDOL Testing Process}
\label{fig:process}
\end{figure}

The IDOL testing process is illustrated in Figure~\ref{fig:process}. First, a series of test programs are generated using the Solsmith tool. Then, a mutation process is applied to these test programs to produce corresponding equivalent variants. 
% The purpose of equivalent mutation is to increase the frequency of compiler optimizations by transforming the program into a form that is more likely to trigger compiler optimizations. 
These equivalent variants next are given as inputs to the Solidity compiler (\emph{i.e.,} solc) and sloc compiles them at different optimization levels. Finally, for each variant, the output results at different optimization levels are compared. If discrepancies are found, it suggests that there is a bug in at least one optimization path. We conducted
a preliminary evaluation of IDOL using 160,000 test programs generated by Solsmith, and have identified three confirmed compiler optimization bugs. As a comparison, we also tried the original DOL approach using the 160,000 generated test programs, but these three bugs are not detected.
\vspace{-15pt}

\textbf{Optimizer Keccak Cache Bug.} For Keccak-256 hash values with the same memory content but different sizes, the compiler incorrectly treats them as equal, causing the bytecode optimizer to mistakenly reuse previously computed hash values.

\textbf{Traditional Code Generation Pipeline Bug.} The cause is that the traditional code pipeline does not evaluate complex expressions, preventing the contract from executing these expressions, which in turn leads to incorrect contract behavior.

\textbf{FullInliner Non-Expression Split Parameter Evaluation Order Bug.} When the compiler performs FullInliner optimization, it fails to correctly handle the splitting of complex expressions, excessively altering the original logic and resulting in an incorrect parameter evaluation order.

\section{Conclusion}
This paper presents an innovative IDOL testing method. The key idea behind IDOL is to perform reverse optimization transformations to generate semantically equivalent variants of the smart contracts under test, aiming to maximize the opportunities to trigger the compiler’s optimization logic. An initial evaluation of IDOL uncovers three optimization bugs, demonstrating the effectiveness and potential of the approach.

%whose key idea behind is to perform reverse transformations on typical compiler optimizations to generate semantically equivalent variants of the smart contracts under test, aiming to maximize the opportunities to trigger
%the compiler’s optimization logic. The method involves firstly series of transformations to the test programs, including loop invariant hoisting, loop reversal, loop condition hoisting, and solc's built-in inverse optimization transformations, converting them into equivalent forms that are more likely to trigger compiler optimizations. These programs are then tested under different optimization levels. The IDOL testing method uncovered three bugs in the Solidity compiler that occur during code optimization. Future work will explore more diverse optimization transformations to further improve the reliability of Solidity compilers.

% improving the reliability of compiler optimization strategies to ensure the security of smart contracts.

\balance 
\bibliographystyle{ieeetr}
\bibliography{references}

\end{document}